\documentclass[aps,prd,twocolumn]{revtex4}
\usepackage{graphicx}
\newcommand{\be}{\begin{equation}}
\newcommand{\ee}{\end{equation}}
\newcommand{\bea}{\begin{eqnarray}}
\newcommand{\eea}{\end{eqnarray}}

\newcommand{\hs}[1]{\hspace{#1 mm}}

\begin{document}


\title{Wrapped Branes and Compact Extra Dimensions in Cosmology}


\author{Ali Kaya}
\email[e-mail:]{kaya@gursey.gov.tr}
\author{Tongu\c{c} Rador}
\email[e-mail:]{rador@gursey.gov.tr}
\affiliation{Feza G\"{u}rsey Institute,\\ Emek Mah. No:68, \c{C}engelk\"{o}y, 81220, \.Istanbul, Turkey}

\date{\today}

\begin{abstract}
We present a cosmological model in $1+m+p$ dimensions, where in
$m$-dimensional space there are uniformly distributed $p$-branes
wrapping over the extra $p$-dimensions. We find that during
cosmological evolution $m$-dimensional space expands with the exact 
power-law corresponding to pressureless matter while the extra
$p$-dimensions contract. Adding matter, we also obtain solutions having
the same property. We show that this might explain in a natural way
why the extra dimensions are small compared to the observed three
spatial directions.  
\end{abstract}
\pacs{}
\maketitle

\section{Introduction}

The inflationary paradigm is successful in explaining away the basic
shortcomings of standard cosmology, like the monopole, horizon and
flatness problems. One might along this line claim that we now have a
cosmological scenario which remains plausible very close to big bang.
However, the possibility of an initial singularity 
still remains to be dealt with. One hopes that this
problem will be resolved once we understand the theory of gravity down
to sizes comparable to Planck length. String theory is one of the
leading candidates for that but for its consistency one has to introduce extra
dimensions. Observationally if these extra dimensions exist their sizes are
much smaller than the size of our perceived universe. Thus it is of
importance to seek for cosmological models where this difference can
be accommodated in a  natural way. 

It is plausible that the sizes of all dimensions  started out
the same, possibly close to Planck length. After various cosmological
eras the perceived universe grew to its size we observe
today. Therefore the problem is to explain how the extra dimensions
remained comparatively small.  

In what follows we propose a toy model, which we believe gives an 
answer to this problem in a natural way. The basic motivation is a
flavor of the idea once exposed by Brandenberder and Vafa \cite{bv}
(see also \cite{bv01} and \cite{bv0}): when a
$p$-brane wrap over the extra $p$-dimensions it resists expansion,
much like a rubber band would if wrapped and glued over the surface of a
balloon. There has been considerable activity in the literature on
similar ideas, see for example the subject of ``brane gas cosmology''
\cite{bg1}-\cite{bg9}. 

Our model is formulated in $1+m+p$-dimensions, where $m$  and  $p$
refer to  the  observed and  compact dimensions respectively, in which
there  is a uniform (with respect to $m$)  distribution   of
$p$-branes   wrapping  over  the   extra  $p$-dimensions. In order to
incorporate this assumption it is enough to take the observed space to
be topologically non-compact. At this point it is apparent
that this approach is different than the brane gas cosmology, where it
is postulated that branes can wrap anywhere. 

In this paper  we focus on time  dependent solutions to Einstein
equations  and  find   exact  expressions   where  possible. In the
earlier work on brane (or string) gas cosmology the interest was mainly
on the thermodynamical aspects and on the incorporation
of T-duality invariance of string theory to cosmology (see for instance
\cite{bv}, \cite{bg1} and \cite{bg2}). For that reason, the problem
was studied in the framework of {\it dilaton} gravity and it was found
that the wrapped branes can prevent cosmological
expansion of the internal space. As we will see, the same can be
achieved in Einstein gravity without invoking T-duality invariance. 

Recently, dynamical aspects of brane gas cosmology in M-theory has
been studied in \cite{val1} and \cite{val2}. Our approach is
technically similar. Namely, we also obtain the energy momentum tensor by
coupling the brane action to the gravity action and assume a uniform
distribution of such branes. However, we concentrate on
pure Einstein gravity rather than M-theory.

The main phenomenon we observed 
is that the existence of wrapped $p$-branes makes the $m$-dimensional
space grow in exactly the same way as pressureless matter would, while
the $p$-dimensional compact space is contracted following a power-law
depending on the numerical value of $p$. From this perspective the
model has predictive power since it gives a definite proportion between the
sizes of the observed and the compact directions. 

We have also observed that adding ordinary matter does not change the
mentioned behavior appreciably except for the case  with negative
pressure. For instance, adding radiation one still finds that the
observed space expands and the compact space contracts. On the other
hand, in case of vacuum domination  both observed and compact spaces
will grow exponentially. More on our line of reasonings is presented
in Section III. 

The organization of the manuscript is as follows: In Section II
we show how to wrap the $p$-branes over extra dimensions, present the
resulting Einstein equations and the aforementioned solution. In Section
III we add ordinary matter to the energy-momentum tensor. In Section
IV we give an estimate for the current size of the internal dimensions
using the solutions. In section V, we compare the relative strengths
of the expansion and contraction forced by $p$-branes. The last
section is devoted to conclusions and possible future extensions of
the model. 

\section{Cosmology of wrapped $p$-branes}\label{II}

Consider a $D$-dimensional space-time which has the following metric
\be\label{met1}
ds^2=-e^{2A}dt^2+e^{2B}dx^idx^i+e^{2C}dy^ady^a,
\ee
where $i=1,..,m$, $a=1,..,p$ and the metric functions $A,B,C$ depend
only on time $t$. Here, $x^i$ is chosen to parameterize the
observed directions which we label as $M_m$ and $y^a$
parameterizes the extra space labeled by $M_p$. We also define
$X^\mu=(t,x^i,y^a)$. We would like to
determine the cosmological evolution in the presence of $p$-branes
wrapping over $M_p$. The dynamics of $p$-branes are
determined by the Polyakov action
\be
-\frac{S_p}{T_p}=\int d\xi^{p+1}\sqrt{-\gamma}\left[\gamma^{\alpha\beta}\partial_\alpha X^\mu\partial_\beta
X^\nu g_{\mu\nu} -p+1\right],\label{pol}
\ee
where $\xi^\alpha=(\tau,\sigma^a)$ are world-volume coordinates, $T_p$ is the
$p$-brane tension, $g_{\mu\nu}$ is the space-time and
$\gamma_{\alpha\beta}$ is the world-volume metrics, and $X^\mu(\xi)$ is
the map from world-volume to space-time. It is easy to show that the
following $p$-brane configuration in (\ref{met1}) is an extremum of the
Polyakov action (\ref{pol})
\bea
t&=&\tau,\nonumber\\
x^i&=&x^i_0,\label{config}\\
y^a&=&\sigma^a,\nonumber
\eea
where $x^i_0$ are constants giving the position of the $p$-brane in
$M_m$. We assume that $y^a$ and $\sigma^a$ coordinates are topologically
$S^1$ so that there is no surface term coming from the variation of
the action (\ref{pol}). We would like to calculate the back reaction of
this configuration on the geometry. For that we couple (\ref{pol}) to
$D$-dimensional Einstein-Hilbert action
\be\label{Eins}
S_E=\frac{1}{\kappa^2}\int d^D X \sqrt{-g}\, R,
\ee
so that
\be \label{Stotal}
S=S_E+S_p.
\ee
The energy momentum tensor for the $p$-brane can be calculated from
(\ref{Stotal}) which gives
\be\label{T1}
-\frac{T^{\mu\nu}}{T_{p}}=\int
d\xi^{p+1}\frac{\sqrt{-\gamma}}{\sqrt{-g}}\gamma^{\alpha\beta}\partial_\alpha
X^\mu\partial_\beta X^\nu\delta(X-X(\xi)).
\ee
In the  orthonormal frame ($e^Adt$, $e^Bdx^i$, $e^Cdy^a$),
(\ref{T1}) corresponding to the $p$-brane configuration (\ref{config})
takes the form
\bea
T_{\hat{t}\hat{t}}&=&T_p e^{-mB}\delta(x-x_0),\nonumber\\
T_{\hat{i}\hat{j}}&=&0,\\
T_{\hat{a}\hat{b}}&=&-T_p e^{-mB}\delta(x-x_0)\delta_{ab}.\nonumber
\eea
For more than one $p$-brane one has to change $\delta(x-x_0)$
with the sum $\sum n_l\delta(x-x_l)$ where $n_l$ is the number of
coincident $p$-branes at the position $x^i_l$. We now assume that
there are uniformly distributed such $p$-branes in $M_m$ and do the
following replacement
\be
\sum n_l\,\delta(x-x_l)\to \int dx'\, n(x')\,\delta(x-x'),
\ee
where $n(x)$ is the number of $p$-branes per unit volume at the
position $x$. Homogeneity of $M_m$ implies that $n(x)$ is constant and
this gives the following energy momentum tensor
\bea
T_{\hat{t}\hat{t}}&=&n\,T_p e^{-mB},\nonumber\\
T_{\hat{i}\hat{j}}&=&0,\label{ebein}\\
T_{\hat{a}\hat{b}}&=&-n\,T_p e^{-mB}\delta_{ab}.\nonumber
\eea
One can easily check  that $\nabla_\mu T^{\mu\nu}=0$. We also note that
a scaling of $x$ coordinates $x\to \lambda x$ requires, by
definition, a scaling of $n$ by  $n\to n/\lambda$.

In solving Einstein equations, we first impose the gauge choice $A=mB+pC$
which fixes $t$-reparameterization invariance in the metric (\ref{met1}).
In this gauge, the field equations that follow from the action
(\ref{Stotal}) can be written as
\bea
A''-A'^2&+&mB'^2+pC'^2=C''\nonumber\\
B''&=&\frac{p+1}{m+p-1}\,\,\kappa^2\,n\,T_p\,\,e^{mB+2pC}\label{E1}\\
C''&=&-\frac{m-2}{m+p-1}\,\,\kappa^2\,n\,T_p\,\,e^{mB+2pC}\nonumber
\eea
where $'$ denotes differentiation with respect to $t$.
The last two equations in (\ref{E1}) imply that $B$ and $C$ are
proportional to each other upto the linear terms in $t$. Ignoring
these terms $B$ and $C$ can be solved up to an
undetermined integration constant. The first equation in (\ref{E1})
then fixes this integration constant. Switching to the proper
time coordinate ({\em which we again denote by $t$}), we finally obtain the
following metric
\be\label{m1}
ds^2=-dt^2+(\alpha t)^{\frac{4}{m}}dx^idx^i + (\alpha t)^{-\frac{4(m-2)}{m(p+1)}}dy^ady^a,
\ee
where
\be
\alpha^2=\frac{m^2(p+1)^2}{2(m+p-1)}\,\frac{\kappa^2\,n\,T_p}{(m-mp+4p)}.
\ee
For the physically important case of $m=3$, from (\ref{m1}), the
scale factors $R_{3}$ and $R_{p}$ of the observed and compact
dimensions respectively are determined as follows 
\bea\label{finres1}
R_{3}(t) &=& (\alpha t)^{2/3}\;,\label{fnref}\\
R_{p}(t) &=& (\alpha t)^{-2/3(p+1)}\label{fn2}\; .
\eea
It is evident that the power-law of the observed space is exactly the
same as the one for pressureless matter in standard cosmology. This is
somewhat expected since the observed space part of energy momentum
tensor vanishes as it does for pressureless matter. On the other hand,
(\ref{fn2}) shows that even in pure Einstein gravity wrapped
branes can prevent expansion of the internal dimensions. 
This is contrary to the general
expectation in the literature (see for instance \cite{bv}, \cite{bg9}
and \cite{val2}) which is based on the intuition that
negative pressure would increase the expansion rate (note that branes
apply negative pressure along the wrapping directions) as in vacuum
domination during inflation. However, (\ref{fn2}) indicates that
negative pressure does not necessarily imply expansion in Einstein
gravity. Moreover, even in de Sitter phase of the early universe,
negative pressure would give exponential contraction with a negative
Hubble constant.    

\section{Adding Matter}\label{addmatter}

We now add ordinary matter to analyze a more realistic model. We will
use for generality the following energy momentum tensor for matter
\be\label{mattert}
T_{\hat{\mu}\hat{\nu}}=\mathrm{diag}(\rho,\,\,p_{\hat{i}},\,\,p_{\hat{a}})\;\;,
\ee
where the indices refer to the obvious orthonormal frame in
(\ref{met1}) and 
\bea
p_{\hat{i}}=\omega\rho,\\
p_{\hat{a}}=\nu\rho,
\eea
where $\omega$ and $\nu$ are constants. The energy-momentum
conservation $\nabla_\mu T^{\mu\nu}=0$ determines $\rho$ in terms
of the metric functions 
\be
\rho\,=\,\rho_0\,e^{-(1+\omega)mB-(1+\nu)pC},
\ee
where $\rho_0$ is a constant and we again impose the gauge $A=mB+pC$.
In the presence of matter, Einstein equations (\ref{E1}) are modified
as follows 
\begin{widetext}
\bea
A''-A'^2+mB'^2+pC'^2&=&C''-\kappa^2\rho_0
(1+\nu)\,e^{(1-\omega)mB+(1-\nu)pC},\nonumber \\ 
(m+p-1) B''&=& \kappa^2 e^{mB+2pC}\left[(p+1)nT_p+\rho_0(1+(p-1)\omega-p\nu)e^{-m\omega B-(1+\nu)pC}\right],\label{E2}\\ 
(m+p-1) C''&=&\kappa^2
e^{mB+2pC}\left[-(m-2)nT_p+\rho_0(1+(m-1)\nu-m\omega)e^{-m\omega B-(1+\nu)pC}\right].\nonumber
\eea
\end{widetext}

These equations have a much richer solution space and it is not
possible in this manuscript to exhaust every interesting one. 
There are various possibilities here depending on what
one chooses for $\omega$ and $\nu$. However it is of crucial importance
to accommodate for the inflationary paradigm, so we consider this first. 
In the 4-dimensional cosmology one chooses $\omega=-1$
to achieve an exponential growth. This however also has a physical
interpretation.  The inflationary solution is the one in which vacuum
energy is dominant. It would be, we feel, quite unnatural to assume
that a constant cosmological constant to be absent from the extra
dimensions, so we belive it is necessary to chose $\nu=-1$ also.  
In this case, comparing the terms on the right hand side of
(\ref{E2}), it is clear that the functions multiplying $\rho_{o}$ is
$e^{mB}$ times larger than the terms coming from $p$-brane sources. 
Evidently the former term will dominate the equations in time granted
the scale $e^B$ is increasing \footnote{Note that both the cosmological
constant and wrapped $p$-branes force $B$ to increase. This
argument would fail if, for instance, we would have the function $C$
instead of $B$ since the cosmological constant and the $p$-branes have
opposite effects on $C$.}. In this regime, $p$-brane sources can be
self-consistently ignored (i.e. one can set $T_p=0$) and the only
contribution to the energy momentum tensor comes from a cosmological
constant, $\rho_0$, which will yield the usual exponential growth of the
inflationary period \footnote{To see this more explicitely, set
$T_p=0$ in (\ref{E2}). Then, it is easy to show that $A=-\ln(Ht)$ and
$B=C=-\ln(Ht)/(m+p)$ is a solution where
$H^2=2\kappa^2\rho_0(m+p)/(m+p-1)$. Making a further coordinate
transformation, one can see that (\ref{met1}) represents de Sitter
space.} where all dimensions expand with the same exponent.

To support the above argument, we also made a simple numerical
integration of the equations (\ref{E2}) in the {\it proper} time
coordinate. We took $p=1$, $m=3$ and $nT_p/\rho_0=10.0$. Also the initial
conditions for the run were such that near $t=0$ the scale factors
obey (\ref{fnref}) and (\ref{fn2}). The resulting plot is presented in
Fig. 1. Note that the inflationary regime sets in around $C''=0$.
Before this, the internal dimensions contract. Consequently, even
though the $e$-foldings after the cosmological constant dominates 
are the same, the internal dimensions exit the inflationary period
with a smaller final value for the scale factor than that of the
observed dimensions. This difference depends on $nT_p/\rho_0$. For our
numerical run, the ratios of the scale factors (i.e. $e^B/e^C$) can be
read from the graph and one finds that the scale factor of the
internal dimensions is about $6.8$ times smaller than that of the
observed dimensions. 

In solving (\ref{E2}) for other cases of $\omega$ and $\nu$, we start
with the following ansatz 
\bea
B=b_1\ln (t)+\ln b_2, \label{ans1}\\
C=c_1\ln (t)+\ln c_2. \label{ans2}
\eea
Using the last two equations in (\ref{E2}) one can uniquely determine these
four constants and the first equation is satisfied
identically. The constants which determine the power-law for
expansion turn out to be
\be\label{b1c1}
b_1=-\frac{2(1+\nu)}{m(1-2\omega+\nu)},\,\,\,\,c_1=\frac{2\omega}{p(1-2\omega+\nu)}
\ee
Form (\ref{ans1}), (\ref{ans2}) and (\ref{b1c1}), one can also fix the
metric function $A$ using the gauge $A=mB+pC$. Thus, all unknown
functions in (\ref{met1}) are now determined. Making a coordinate
transformation to switch to the proper time coordinate ({\em which we
again denote by $t$}), the metric can be written as
\be\label{met2}
ds^2\,=\,-dt^2\, + \,(\alpha_1 t)^{\frac{4}{m}}dx^idx^i\,+\,(\alpha_2
t)^{\frac{-4\omega}{p(1+\nu)}}dy^ady^a,
\ee
where $\alpha_1$ and $\alpha_2$ depend on $b_2$ and $c_2$, and they  
can be set to 1 by scalings of $x$ and $y$ coordinates.  

\begin{figure}
\centerline{
\includegraphics[width=8.0cm]{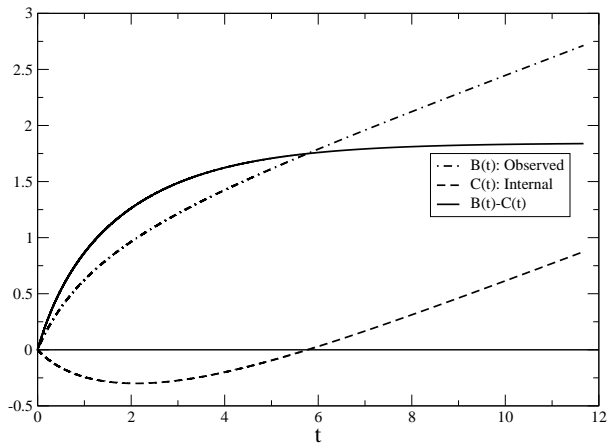}}
\caption{Evolution of the scale factors $B$ and $C$, for $\omega=\nu=-1$,
$p=1$, $m=3$ and $nT_p/\rho_0=10.0$. See text for the initial conditions.}
\end{figure}

It is clear from (\ref{met2}) that one should choose $\nu\not=-1$. 
Also, for the above ansatz to work, $b_2$ and $c_2$ should be
determined to be positive numbers. This imposes, after a
straightforward but somewhat tedious algebra,  the following
conditions on the parameters of the model
\bea
\hs{-4}\frac{(1+\nu)(1-\nu)}{m}+\frac{\omega(1-\omega)}{p}&>&(1-\omega+\nu)(\omega-\nu)\label{ineq1}\\
\frac{(1+\nu)(m-2)}{m}&>&\frac{\omega(p+1)}{p}\label{ineq2}\\
1-2\omega+\nu&\not=&0\label{ineq3}.
\eea

For pressureless dust $\omega=\nu=0$, and these conditions are
satisfied for all $m>2$ and $p$. In this case the scale
factor for the extra dimensions is equal to 1; the expansion  
forced by the dust is compensated by the contraction forced by
$p$-branes. 

For radiation, two different equations of state are possible. When the
sizes of all dimensions are close to each other one has
$\omega=\nu=1/(m+p)$. In this case, (\ref{ineq1}) and (\ref{ineq3})
are satisfied identically. However, (\ref{ineq2}) restricts the
possible values of $m$ and $p$. For instance, for $m=3$ (\ref{ineq2}) imposes
that $p>1$. On the other hand, when the extra dimensions are very
small compared to the observed dimensions one has $\omega=1/m$ and
$\nu=0$. In this case (\ref{ineq1}) and (\ref{ineq3}) are satisfied
identically, but (\ref{ineq2}) implies that $mp>3p+1$ and thus one
should take $m>4$. 

Before closing this section let us emphasize that these constraints
only show that the ansatz chosen in solving the differential
equations is not valid for all parameters. However, one can still
infer a general dynamical behavior from the above solutions. We
belive that for the cases where (\ref{met2}) is not valid, one would 
observe similar effects for ordinary matter coupled to $p$-branes. 
 
\section{An estimate for the size of the internal dimensions}\label{scenario}

Here we try to obtain an estimate for the current size of the extra
dimensions along the lines of the model we presented. In this section
we set $m=3$.  We take the universe to be filled with ordinary matter
(characterized with an equation of state depending on $\omega,\nu$)
and wrapped branes. We further assume that after big bang all dimensions
started out close to Planck length. The standard model of cosmology
tells us that the universe passed through three different eras. First, an
inflationary period took place. After inflation, there was a radiation
dominated era followed by a matter dominated one  which still 
(possibly \footnote{Recent observations indicate that the universe is
accelerating which suggests that the energy density is now dominated
with some kind of undetermined energy (dark energy) having negative
pressure. In  this paper we are not going to discuss possible
modifications implied by the existence of dark energy to our scenario.})
is going on.  We assume that adding wrapped $p$-branes along the extra
dimensions do not alter this history (but modify the power-law
expansion as we discussed). 

During inflation, one should take $\omega=\nu=-1$ which simply
represents a positive cosmological constant. As we discussed in the
previous section (\ref{E2}) gives the usual exponential
growth (with a quantitative modification, see Fig. 1). Assuming
approximately $70$ e-foldings during inflation (which is required by
cosmological phenomenology), the sizes of {\it all} dimensions grew to
about $10^{-5}$m from the Planck length. Note that we are not
attempting to answer questions like what is deriving inflation or how
the graceful exit occurs.   

Just after the inflation during radiation dominated era one can set
$\omega=\nu=1/(p+3)$. Note that the temperature at the beginning
of the radiation era is expected to be of the order of $10^{15}$K
which corresponds to a length scale $10^{-18}$m. Since this is much
smaller than $10^{-5}$m, the radiation is allowed to apply
pressure along the extra dimensions. In this period, the solution
(\ref{met2}) can be used to describe the cosmological evolution of the
universe \footnote{Since in (\ref{met2}) 
the internal space contracts,  the equation of
state for radiation also starts to shift. As the extra
dimensions become much smaller than the scale set by the temperature,
one should set $\omega =1/3$ and $\nu=0$. We ignore this subtlety
in the following since we try to obtain a rough estimate to see if the
scenario makes sense in a first crude approximation.}. 
Assuming  that the vacuum energy driving the inflation
was about $10^{14}$ GeV, the Hubble parameter of the inflation can be 
determined to be $H^{-1}\approx 10^{-35}$ sec. We match the solution
(\ref{met2}) to de Sitter space by demanding that the
Hubble parameter of the observed universe is continuous. Therefore, 
(\ref{met2}) is valid starting from $t\approx 10^{-35}$
sec. Using that the radiation domination ended at about $10^{11}$ sec,
(\ref{met2}) gives the size of the internal space at the end of
radiation to be $10^{-5}\times 10^{-138/[p(p+4)]}$m.

As the observed universe expands, pressureless matter started to
dominate the cosmological evolution and one can then set
$\omega=\nu=0$. During this period, (\ref{met2}) implies that the
extra dimensions stay fixed. 

Summarizing, we found  that after an exponential growth during inflation,
the extra dimensions contracted till the end of radiation dominated
era and then remained unaltered. This gives the following estimate for
the present size ($r_{p}$) of the extra dimensions  
\be
r_{p}\approx 10^{-5}\times10^{-\frac{138}{p(p+4)}}\;\;\rm
m.\label{predict1}
\ee
For $p=1$ we get $r_{1} \approx 10^{-33}$ m. For higher values
of $p$ one finds much larger estimates. However let us 
remind the reader of the digression we had on the structure of the
inflationary scenarios. In reality, during the inflationary period the
scale factor of the  internal dimensions will be subject to a lesser
growth  than that of the observed dimensions. For instance, for the
numerical case we have considered, $r_p$ should be scaled down by a
factor of $6.8$. Recall that this is for $nT_p/\rho_0=10.0$. Larger
values for that ratio will make $r_p$ smaller.  

\section{More on $p$-brane cosmology}

In section \ref{II}, we found that in the presence of uniformly
distributed wrapped $p$-branes, the observed space expands while the
compact space contracts. In this section, we would like to compare
relative strengths of these two effects by assuming existence of two
branes where they will be uniformly distributed wherever they don't
wrap. Let us consider the following metric in $1+m+p+q$-dimensions  
\bea\label{met3}
ds^2&=&-e^{2A}dt^2+e^{2B}dx^idx^i+e^{2C}dy^{a_{1}}dy^{a_{1}}\nonumber\\
&+&e^{2D}dz^{a{_2}}dz^{a_{2}},
\eea
where $a_{1},b_{1}=1,...,p$, $a_{2},b_{2}=1,...,q$ and the metric
functions depend only on time $t$.  We assume that in (\ref{met3})
there are $p$ and $q$ dimensional branes wrapping over $y$ and $z$
coordinates, respectively. Following our reasonings in section
\ref{II}, one can easily write down the energy momentum tensor
corresponding to this configuration 
\bea
T_{\hat{t}\hat{t}}&=&T_p e^{-mB-qD}+T_qe^{-mB-pC},\nonumber\\
T_{\hat{i}\hat{j}}&=&0,\nonumber\\
T_{\hat{a}_{1}\hat{b}_{1}}&=&-T_p e^{-mB-qD},\\
T_{\hat{a}_{2}\hat{b}_{2}}&=&-T_q e^{-mB-pC},\nonumber
\eea
where $T_p$ and $T_q$ are respective brane tensions and hatted indices
refer to the orthonormal basis in (\ref{met3}). At this point one
can check that $\nabla_\mu T^{\mu\nu}=0$. Imposing the
gauge $A=mB+pC+qD$ the Einstein equations can be written as
\begin{widetext}
\bea
A''-A'^2+mB'^2+pC'^2+qD'^2&=&C''+D''-B'',\nonumber \\ 
(m+p+q-1)B''&=&\kappa^2 e^{mB+pC+qD}\left[(p+1)n_pT_p
  e^{pC}+(q+1)n_qT_q e^{qD}\right],\nonumber\\ 
(m+p+q-1)C''&=&\kappa^2 e^{mB+pC+qD}\left[-(m+q-2)n_pT_p
  e^{pC}+(q+1)n_qT_q e^{qD}\right],\label{E3}\\ 
(m+p+q-1)D''&=&\kappa^2 e^{mB+pC+qD}\left[(p+1)n_pT_p
  e^{pC}-(m+p-2)n_qT_q e^{qD}\right],\nonumber
\eea
\end{widetext}
where $n_p$ and $n_q$ are number of branes per co-moving volumes
parameterized by $(x,z)$ and $(x,y)$ coordinates respectively. To solve
these equations exactly we first note that the last three equations in
(\ref{E3}) imply that 
\be\label{35}
(m-3)B+(p+1)C+(q+1)D=0
\ee
up to the terms linear in $t$ which we ignore in the following.
Using (\ref{35}), the middle two equations can be solely expressed
in terms of $B$ and $C$. To find a power-law solution we choose 
\bea
B=b_1\ln (t)+\ln b_2, \\
C=c_1\ln (t)+\ln c_2. 
\eea
It is now straightforward to check that the four constants can be
determined uniquely. For instance we find
\bea
b_1&=&\frac{-2(p+q+2pq)}{9pq+m(p+q-pq)},\\
c_1&=&\frac{2(m-3)q}{9pq+m(p+q-pq)}.
\eea
Using (\ref{35}) and the gauge choice $A=mB+pC+qD$, one can also
determine metric functions $A$ and $D$. The first equation in
(\ref{E3}) is then to be checked for consistency. A straightforward
calculation shows that it is satisfied identically. Let us also
emphasize that, contrary to the case we encountered in section III,
the constants $b_2$ and $c_2$ turn out to be positive for all possible
values of $m$, $p$ and $q$.  

Using the above results, in the {\it proper time coordinate} the metric can
be written as   
\be
ds^2=-dt^2+ R_m^2 dx^idx^i + R_p^2 dy^{a_1}dy^{a_1}+R_q^2 dz^{a_2}dy^{a_2}
\ee
where
\bea
\ln(R_m)&=&{\frac{2(p+q+2pq)}{3pq+m(p+q+pq)}}\;\ln(\alpha_m t)\nonumber\\
\ln(R_p)&=&-{\frac{2q(m-3)}{3pq+m(p+q+pq)}}\;\ln(\alpha_p t)\label{ek1}\\
\ln(R_q)&=&-{\frac{2p(m-3)}{3pq+m(p+q+pq)}}\;\ln(\alpha_q t)\nonumber
\eea
and $\alpha_m$, $\alpha_p$ and $\alpha_q$ are positive constants. 
Comparing (\ref{ek1}) with the results of section II, we see
that wrapping a $p$-brane and a $q$-brane is physically distinct from
wrapping a $p+q$-brane. The main difference is that in the former case
the $p$-branes are uniformly distributed along the dimensions
over which $q$-branes wrapped  and vice versa. 

Another interesting aspect is that {\it only} the dimension of the
observed space $m$ determines whether the compact dimensions expand or
contract. For the two brane configuration considered above we see that
$m=3$ is the critical value where the internal dimensions are
stabilized by the cancellation of the expansion forced by the uniformly
distributed $p$-branes with the contraction forced by the wrapping of
$q$-branes. We find that the compact dimensions expand for $m>3$ and
they diminish for $m<3$. 

Generalizing (\ref{m1}) and (\ref{ek1}), one would guess that the exponent of
the power-law of the internal dimensions is proportional to $m-k-1$
where $k$ is the number of partitionings of the extra dimensions, i.e
number of distinct brane configurations. For large $k$, one should
increase $m$ to have contracting extra dimensions.   

\section{Conclusions  and future directions}{\label{conclusions}}

We have shown that if one allows for $p$-branes wrapping around extra
dimensions Einstein equations allow for solutions where their size
diminishes during cosmological evolution. In this paper, we have
focused on the main aspects of the idea and omitted various details
which may form topics for future studies.

First, since the $p$-branes allow for power-law solutions for
the observed universe which are exactly the same as pressureless matter
one can think of them as a form of dark matter. 

Second, it is an important question to ask about
the time of forming of wrapped $p$-branes during cosmological
evolution. In this work we have assumed that they did exist since the
beginning of time but it is plausible that they may form later. One
important question along this line is about the connection of the
wrapped branes to the exit from inflation. 


Let us also recall that Brandenberger and Vafa argued that
\cite{bv} string interactions may yield an upper bound ($m\leq3$) on
the number of observed dimensions if one assumes that all the
directions are compact. We note in this context that
the two brane scenario we have discussed in section V yields a lower
bound for $m$ if one insists on preventing expansion of the extra
dimensions during entire cosmological history. One is tempted to
speculate that the two ideas might be merged to fix $m$. 

Finally, one could further think of adding  ordinary matter to the
configuration discussed in section V. This is  expected to make
the internal dimensions grow for $m\geq 3$. Of course a detailed
numerical analysis might prove otherwise. One further extension could
be to wrap branes over branes. In this case we expect the effects to
add up in the directions the branes intersect. And this might result
in a hierarchy of scale factors.


\begin{thebibliography}{20}
\bibitem{bv} Branderberger R. and Vafa C., Nucl. Phys. {\bf B316},
  391,(1989).
\bibitem{bv01} Nishimura H. and Tabuse M., Mod. Phys. Lett. {\bf A2},
  299, (1987).
\bibitem{bv0} Kripfganz J. and Perlt H.,
 Class. Quant. Grav. {\bf 5}, 453, (1988). 
\bibitem{bg1} Tseytlin A. A. and Vafa C., Nucl. Phys. {\bf B 372}, 443, (1992).
\bibitem{bg2} Tseytlin A. A., Class. Quant. Grav. {\bf 9}, 979, (1992).
\bibitem{bg3} Cleaver G. B. and Rosenthal P. J., Nucl. Phys. {\bf B
457}, 621, (1995).
\bibitem{bg4} Park C., Sink S. and Lee S., Phys. Rev. {\bf D 61},
083514, (2000).
\bibitem{bg5} Alexander S., Brandenberger R. H. and Easson D.,
  Phys. Rev. {\bf D 62}, 103509, (2000).
\bibitem{bgx} Easson, D.A., hep-th/0110225.
\bibitem{bgxx} Easson, D.A., hep-th/0111055.
\bibitem{bg6} Brandenberger R. H., Easson D. A. and Kimberly D.,
  Nucl. Phys. {\bf B 623}, 421, (2002).
\bibitem{bg7} Easther R., Greene B. R., and Jackson M. G.,
  Phys. Rev. {\bf D 66}, 023502, (2002).
\bibitem{bg8} Watson S., Brandenberger R. H., hep-th/0207168.
\bibitem{bg9} Boehm T. and Branderberger R., hep-th/0208188.
\bibitem{val1} Easther R., Greene B. R., Jackson
  M. G. and Kabat D., hep-th/0211124.
\bibitem{val2} Alexander S. H. S., hep-th/0212151. 
\end{thebibliography}
\end{document}